\documentclass[fleqn,10pt,english]{SelfArx}
\usepackage{lipsum}

\definecolor{color1}{RGB}{0,0,90} 
\definecolor{color2}{RGB}{0,20,20} 

\usepackage{hyperref} 
\hypersetup{hidelinks,colorlinks,breaklinks=true,urlcolor=color2,citecolor=color1,linkcolor=color1,bookmarksopen=false,pdftitle={Title},pdfauthor={Author}}

 \renewcommand{\vec}[1]{\mbox{\boldmath $#1$}}

 \def\gsim{\lower.4ex\hbox{$\;\buildrel >\over{\scriptstyle\sim}\;$}}
 \def\lsim{\lower.4ex\hbox{$\;\buildrel <\over{\scriptstyle\sim}\;$}}

 \def\aap{A\&A}
 \def\apj{ApJ}
 
 \def\apjs{ApJS}
 \def\an{AN}

 \def\sp{SoPh}

\JournalInfo{Astronomy Letters, 2019, Vol. 45, No. 1} 
\Archive{}
\PaperTitle{
Large-Scale Magnetic Field Fragmentation in Flux-Tubes
Near the Base of the Solar Convection Zone
}
\Authors{L.~L.~Kitchatinov*}
\affiliation{\textit{Institute of Solar-Terrestrial Physics, Lermontov Str. 126A, Irkutsk, 664033, Russian Federation}}
\affiliation{*\textbf{E-mail}: kit@iszf.irk.ru}

\newcommand{\keywordname}{Keywords}

\Abstract{Magnetic quenching of turbulent thermal diffusivity leads to instability of the large-scale field with the production of spatially isolated regions of enhanced field. This conclusion follows from a linear stability analysis in the framework of mean-field magnetohydrodynamics that allows for thermal diffusivity dependence on the magnetic field. The characteristic growth time of the instability is short compared to the 11-year period of solar activity. The characteristic scale of the increased field regions measures in tens of mega-meters. The instability can produce magnetic inhomogeneities whose buoyant rise to the solar surface forms the solar active regions. The magnetic energy of the field concentrations coincides in order of magnitude with the energy of the active regions.}

\begin{document}

\flushbottom 

\maketitle 


\thispagestyle{empty} 

\section{Introduction} 
The global magnetic field of the Sun and relatively small-scale fields of its active regions are mutually related. Babcock (1961) was probably the first to note that Joy's law (Hale et al. 1919) for sunspot groups can be the reason for the 11-year cyclic variations of the global field. Estimations based on sunspot data support the operation of the mechanism envisaged by Babcock on the Sun (Erofeev 2004; Dasi-Espuig et al. 2010; Kitchatinov \& Olemskoy 2011). On the other hand, magnetic fields of the active regions can be related to the emergence of global toroidal field fragments -- flux-tubes -- to the solar surface\footnote{The term \lq flux-tube' is used here as a short name for the regions of enhanced mainly azimuthal magnetic field, not for ideal tubes of constant circular cross-section.}. This picture is supported by observations of the active regions (see, e.g., Zwaan 1992; Lites et al. 1998; Khlystova \& Toriumi 2017). Modeling of buoyant flux-tubes
reproduces the Joy's law for the active regions (D'Silva \& Choudhuri 1993).

The problem however is that rising flux-tube models agree with observations for sufficiently strong fields of the order of $10^5$\,G only (D'Silva \& Choudhuri 1993; Caligari et al. 1995; Weber et al. 2011). More specifically, the fields should have this strength near the base of the convection zone from where they start their rise to the solar surface. Solar dynamo models do not show fields of such strength. This is natural: convective dynamos cannot amplify the fields to an energy density exceeding the kinetic energy of the field generating flows. The equipartition field,
\begin{equation}
    B_\mathrm{eq} = \sqrt{4\pi\rho}\ u
    \label{1}
\end{equation}
($\rho$ is density and $u$ is the {\sl rms} convective velocity), reaches its maximum strength $\lsim 10^4$\,G near the base of the solar convection zone. The fact that most successful current models for the solar dynamo are based on the Babcock-Leighton mechanism (Babcock 1961; Leighton 1969) and therefore implicitly assume strong near-base fields redoubles the problem.

This seeming contradiction is usually sidestepped with the assumption that the mean dynamo-field of several kilo-Gauss consists of isolated flux-tubes of much stronger fields. The mechanism capable of amplifying the field to the strength of $\sim$100\,kG was not specified, however. Isolated regions of relatively strong fields can result from magnetic buoyancy instability (Parker 1979) or can be due to the field expulsion from the regions of circular motion (Weiss 1966), however, these do not give an amplification above the equipartition level of Eq.\,(\ref{1}). To exceed this level, a more powerful source of energy than the energy of convective motions is required, e.g., thermal energy. A promising possibility was noticed by Parker (1984): isolated regions of a strong field can result from magnetic suppression of convective heat transport.

Parker's idea was as follows. Magnetic field suppresses convection. This leads to an increase in super-adiabatic gradient and, therefore, to an increase in thermal energy in the convection zone. Spatial redistribution of the field with its intermittent concentration in  flux-tubes can be \lq energetically profitable'. An increase in magnetic energy from such a redistribution can be overcompensated by a decrease in thermal energy due to amplified convective heat transport in the weak field regions between the tubes.

This paper makes a first step in the quantitative analysis of this possibility. It concerns a layer with a horizontal magnetic field, which mimics the large-scale toroidal field of the Sun, near the base of the convection zone. The mean-field approach is applied, i.e. convection is accounted for implicitly by introducing effective (turbulent) transport coefficients. The effective thermal conductivity depends on the strength of the magnetic field. A similar approach was formerly applied to the problem of sunspot equilibrium (Kitchatinov \& Mazur 2000; Kitchatinov \& Olemskoy 2006). In the absence of the magnetic field, the layer is stable (in the framework of the mean-field approach, as already stated). In the presence of the magnetic field, instability takes place producing isolated regions of the enhanced field. The paper is confined to the linear stability analysis. The amplitude of the field \lq bunches' remains therefore uncertain. Computations, however, show that the increase in thermal energy due to magnetic quenching of thermal diffusivity exceeds the magnetic energy. The field amplification at the nonlinear stage of the instability can, therefore, be large.
\section{Problem formulation}
\subsection{Main parameters and design of the model}
We consider a horizontal layer of thickness $h$ near the base of the convection zone, where the solar dynamo is expected to produce its strongest fields. Our approach demands that the layer be entirely embedded in the convection zone. The bottom boundary is however placed as close as possible to the base of this zone. The density $\rho_0 = 0.15$\,g/cm$^3$, temperature $T_0 = 2.1\times 10^6$\,K, and gravity $g = 5\times 10^5$\,cm$^2$/s at the bottom boundary are therefore taken from the solar structure model for the heliocentric distance $r_\mathrm{b}$ where the radiative heat flux
\begin{equation}
    F^\mathrm{rad} = - \frac{16\sigma T^3}{3\kappa\rho}
    \frac{\partial T}{\partial r}
    \label{2}
\end{equation}
is only marginally smaller than the total flux: $F^\mathrm{rad} = (1 -
\varepsilon)L_\odot/(4\pi r_\mathrm{b}^2)$, $\varepsilon \simeq 10^{-3}$ (cf. Stix 1989). In Eq.\,(\ref{2}), $\sigma$ is the Stefan-Boltzmann constant, $\kappa$ is the opacity, and other standard notations are used. The opacity is computed with the {\sl OPAL} tables\footnote{https://opalopacity.llnl.gov}. The relative (by mass) content of hydrogen and heavy elements in this computations where taken $X = 0.71$ and $Z = 0.02$ respectively.

Spherical curvature is neglected and the plane layer is unbounded in horizontal dimensions. The Cartesian coordinate system is used with its $z=0$ plane being the bottom boundary of the layer. The $z$-axis points upward.

The relative deviation of the density and temperature gradients from their adiabatic values in the depths of the convection zone are small ($\lsim 10^{-5}$). The lower part of the convection zone \lq\lq lies on essentially the same adiabate'' (Gilman 1986, p.98). Deviations from the adiabatic profiles,
\begin{eqnarray}
    T(z) &=& T_0\left( 1 - z/H\right),\ \ H = c_\mathrm{p} T_0/g,
    \nonumber \\
    \rho(z) &=& \rho_0\left(1 - z/H\right)^{\frac{1}{\gamma - 1}},
    \label{3}
\end{eqnarray}
are therefore neglected. In this equation, $c_\mathrm{p} = 3.45\times 10^8$\,cgs is the specific heat at constant pressure and $\gamma = c_\mathrm{p}/c_\mathrm{v} = 5/3$ is the adiabaticity index. The deviation from adiabaticity cannot be neglected, however, in the specific entropy $S = c_\mathrm{v}\mathrm{ln}(P/\rho^\gamma)$ ($P$ is the pressure) whose gradient is not small compared to the (zero) gradient for the adiabatic stratification.

The constant heat flux $F = L_\odot/(4\pi r_\mathrm{b}^2) =  1.226\times
10^{11}$\,erg/(cm$^2$s) enters the layer through its bottom. Inside the layer, the energy is transported by radiation and convection.
\subsection{Equation system and background equilibrium}
As already mentioned, the effect of turbulent convection in the mean-field approach applied is parameterized by the turbulent transport coefficients. The characteristic scale of turbulent convection near the base of the convection zone is however not small compared to the mean-fields scale. In this case, turbulent transport should be described with non-local (integral) equations. The non-local transport theory is still lacking, however, and the local diffusion approximation is used in this paper in the absence of better possibilities.

The magnetic field decreases the transport coefficients and induces their anisotropy: the transport coefficients for the directions along and across the field lines differ. This paper neglects the anisotropy and quenching of the viscosity and magnetic diffusivity. Multiple simplifications and approximations are unavoidable in the complicated problem. Otherwise, the physics of the results are difficult to interpret.

The expected flux-tube formation is related to magnetic quenching of the thermal diffusivity. The heat transport equation
\begin{equation}
    \rho T \left(\frac{\partial S}{\partial t}
    + {\vec u}\cdot{\vec\nabla} S\right)\ =
    {\vec\nabla}\cdot\left(\rho T\chi{\vec\nabla} S
    - {\vec F}^\mathrm{rad}\right) ,
    \label{4}
\end{equation}
therefore, keeps the dependence of turbulent diffusivity $\chi$ on the magnetic field:
\begin{equation}
    \chi = \chi_{_\mathrm{T}}\phi (\beta),
    \label{5}
\end{equation}
where $\chi_{_\mathrm{T}}$ is the thermal diffusivity in the absence of the magnetic field and $\beta = B/B_\mathrm{eq}$ is the ratio of the field strength to its equipartition value
(\ref{1}). The function
\begin{equation}
    \phi(\beta) = \frac{3}{8\beta^2}\left( \frac{\beta^2 -1}{\beta^2 + 1}
    + \frac{\beta^2 +1}{\beta}\mathrm{arctg}(\beta )\right)\
    \label{6}
\end{equation}
for the dependence is taken from the quasi-linear theory (Kit\-cha\-ti\-nov et al. 1994).

The relation $u^2 =-\frac{\ell^2 g}{4 c_\mathrm{p}}\frac{\partial S}{\partial z}$ of the mixing-length theory is used to estimate the thermal diffusivity  $\chi_{_\mathrm{T}} = \ell u/3$ for the non-magnetic case ($\ell = \alpha_{_\mathrm{MLT}} H_\mathrm{p}$ is the mixing-length proportional to the pressure scale height $H_\mathrm{p}
=-P\left(\frac{\mathrm{d}P}{\mathrm{d}z}\right)^{-1}$). The steady solution of the equation
(\ref{4}) then gives
\begin{eqnarray}
    \chi_{_\mathrm{T}} &=& \alpha_{_\mathrm{MLT}}^{4/3}
    \frac{(c_\mathrm{p} - c_\mathrm{v})T}{g}
    \left(\frac{(\gamma - 1)\delta F}{36\gamma\rho}\right)^{1/3} ,
    \nonumber \\[0.1 truecm]
    B_\mathrm{eq} &=& \sqrt{\pi}\rho^{1/6}
    \left( 6\alpha_{_\mathrm{MLT}}\frac{\gamma - 1}{\gamma}\delta F\right)^{1/3},
    \label{7}
\end{eqnarray}
where $\delta F = F - F^\mathrm{rad}$ is the convective heat flux.

\begin{figure}[]\centering
\includegraphics[width=\linewidth]{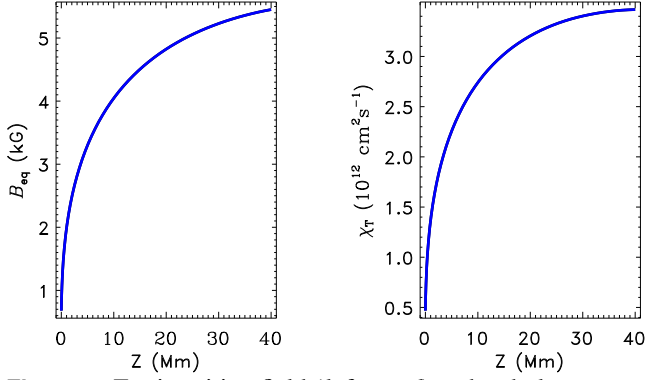}
 \caption{Equipartition field ({\sl left panel})
       and turbulent diffusivity ({\sl right}) versus height $z$ above the base of the convection zone.
    }
 \label{f1}
\end{figure}

Figure\,1 shows the dependence of $\chi_{_\mathrm{T}}$ and $B_\mathrm{eq}$ of Eq.\,(\ref{7}) on the height $z$ above the bottom of the convection zone for the case of the layer thickness $h = 40$\,Mm and $\alpha_{_\mathrm{MLT}} = 0.49$. At the height of 40\,Mm, the diffusivity reaches its maximum value and decreases both downwards and upwards from this height. The choice of the $\alpha_{_\mathrm{MLT}}$-value will be explained later.

The induction equation for the large-scale field,
\begin{equation}
    \frac{\partial{\vec B}}{\partial t} =
    {\vec\nabla}\times\left({\vec v}\times{\vec B}
    - \sqrt{\eta_{_\mathrm{T}}}
    {\vec\nabla}\times(\sqrt{\eta_{_\mathrm{T}}}{\vec B})\right) ,
    \label{8}
\end{equation}
accounts for the diamagnetic pumping with the effective velocity ${\vec v}_\mathrm{dia} = -{\vec\nabla}\eta_{_\mathrm{T}}/2$ (Zeldovich 1957; Krause and R\"adler 1980). The motion equation with turbulent viscosity $\nu_{_\mathrm{T}}$ reads
\begin{eqnarray}
    \rho\frac{\partial v_i}{\partial t} &+& \rho v_j\nabla_jv_i =
    \frac{1}{4\pi}\nabla_j\left(B_iB_j - \delta_{ij}B^2/2\right)
    \nonumber \\
    &+& \nabla_j\rho\nu_{_\mathrm{T}}
    \left(\nabla_jv_i + \nabla_iv_j
    - \frac{2}{3}\delta_{ij}({\vec\nabla}\cdot{\vec v})\right)
    \nonumber \\
    &-& \nabla_iP + \rho g_i,
    \label{9}
\end{eqnarray}
where repetition of subscripts signifies summation.

The Prandtl number Pr\,=\,$\nu_{_\mathrm{T}}/\chi_{_\mathrm{T}}$ and the magnetic Prandtl number Pm\,=\,$\nu_{_\mathrm{T}}/\eta_{_\mathrm{T}}$ for turbulent convection are of order one (Yousef et al. 2003). The quasi-linear theory gives equal values Pr\,=\,Pm\,=0.8 for these numbers (Kitchatinov et al. 1994), which are used in what follows.

The magnetic field in the steady background state of the stability analysis is prescribed with its given value on the top boundary. The field strength $B_0$ on the top boundary is a parameter of the model. As rotation is not included, all horizontal directions are equivalent. The background field ${\vec B}$ is assumed to point along the $y$-axis.  Equations (\ref{8}) and (\ref{4}) give the distributions of the magnetic field and the entropy gradient for the background state,
\begin{eqnarray}
    B(z) &=& B_0\left(\frac{\eta_{_\mathrm{T}}(h)}
    {\eta_{_\mathrm{T}}(z)}\right)^{1/2} ,
    \nonumber \\
    \frac{\partial S_0}{\partial z} &=& -\frac{\delta F}
    {\rho T\chi_{_\mathrm{T}}\phi(\beta)} ,
    \label{10}
\end{eqnarray}
and Eq.\,(\ref{9}) allows the mean flow to be absent, ${\vec v} = 0$. Only the gradient of the entropy - not the entropy value - is required for the stability analysis. Nevertheless, we fix the value with the boundary condition $S(h) = 0$.

\begin{figure}[]
\includegraphics[width=8 truecm]{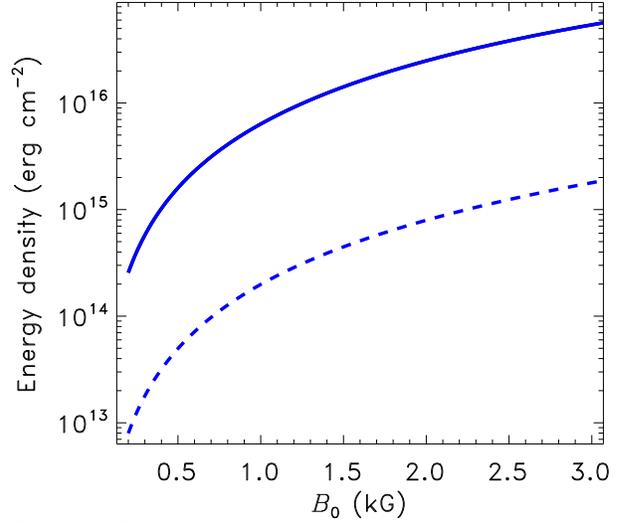}
 \caption{Dependence of the magnetic energy (\ref{11}) ({\sl dashed line})
        and increment in the thermal energy (\ref{12}) ({\sl full line}) on the field strength $B_0$ on the top boundary.
    }
 \label{f2}
\end{figure}

It is remarkable that the surface density of magnetic energy, i.e. the height-integrated energy density
\begin{equation}
    W_\mathrm{B} = \frac{1}{8\pi}\int\limits_0^h\ B^2(z)\mathrm{d}z,
    \label{11}
\end{equation}
is smaller than the magnetically induced increment in the thermal energy
\begin{equation}
    W_\mathrm{T} = \int\limits_0^h\ \rho T \delta S \mathrm{d}z .
    \label{12}
\end{equation}
In this equation, $\delta S$ is the increment in the specific entropy induced by the magnetic field, i.e. the difference in the values of $S$ between the cases of finite and zero magnetic fields. Dependencies of the energies of Eqs. (\ref{11}) and (\ref{12}) on the field strength $B_0$ at the top boundary are shown in Fig.\,2. The increment in thermal energy is about ten times larger than the magnetic energy. Therefore, the field fragmentation in flux-tubes (Parker 1984) can indeed be \lq energetically profitable'.
\subsection{Linear stability problem}
The continuity equation for the depths of the solar convection zone can be written in the inelastic approximation (see, e.g., Gilman \& Glatzmeier 1981), ${\vec\nabla}\cdot (\rho{\vec v}) = 0$. For analysis of the stability of the above-defined equilibrium to small disturbances, it is convenient to split the disturbances of the magnetic field $\vec b$
and momentum density $\rho{\vec v}$ into their toroidal and poloidal parts:
\begin{eqnarray}
    {\vec b} &=& {\vec\nabla}\times\left( \hat{\vec z}T'
    + {\vec\nabla}\times(\hat{\vec z}P')\right),
    \nonumber \\
    {\vec v} &=& \frac{1}{\rho}{\vec\nabla}\times\left( \hat{\vec z}W
    + {\vec\nabla}\times(\hat{\vec z}V)\right),
    \label{13}
\end{eqnarray}
where $\hat{\vec z}$ is the unit vector along the $z$-axis and dashes in the notations for the toroidal ($T'$) and poloidal ($P'$) field potentials distinguish them from the notations for temperature and pressure. Equations (\ref{13}) are introduced by analogy with the stability problems in spherical geometry (Chandrasekhar 1961, p.622) to ensure the divergence-free of the magnetic and flow disturbances.

Linearization of equations (\ref{4}), (\ref{8}) and (\ref{9}) in small disturbances gives a system of five equations for the linear stability problem: four equations for the poloidal and toroidal components of the magnetic field and flow and an equation for the entropy disturbances. Coefficients in these equations do not depend on $x$ and $y$. The wave-type
dependence $\mathrm{exp}(\mathrm{i}k_1x + \mathrm{i}k_2 y)$ on these coordinates can,
therefore, be prescribed.

The linearized entropy equation,
\begin{eqnarray}
    \frac{\partial S}{\partial t} &=& \frac{\mathrm{i}}
    {\rho T}\frac{\partial}{\partial z}
    \left(
    \rho T \phi'(\beta) \frac{\chi_{_\mathrm{T}}}{B_\mathrm{eq}}
    \frac{\mathrm{d}S_0}{\mathrm{d}z}
    \left(
    k_2\frac{\partial P'}{\partial z} - k_1T'
    \right)
    \right)
    \nonumber \\
    &+& \frac{1}{\rho T}\frac{\partial}{\partial z}
    \left(
    \rho T \chi_{_\mathrm{T}}\phi(\beta)\frac{\partial S}{\partial z}
    \right)
    \nonumber \\
    &-& k^2\chi_{_\mathrm{T}}\phi (\beta ) S
    - \frac{k^2}{\rho}\frac{\partial S_0}{\partial z}\ V ,
    \label{14}
\end{eqnarray}
includes the contribution of magnetic disturbances (the first term on the right-hand side). This contribution distinguishes the problem at hand from the standard
analysis of thermal convection. This new contribution comes from the dependence of the effective thermal diffusivity (\ref{5}) on the magnetic field. It affects the solution of the problem considerably. In the Eq.\,(\ref{14}), $k^2 = k_1^2 + k_2^2$ and $\phi '$ signifies the derivative of the function
(\ref{6}).

The divergence-free of the flow (\ref{13}) demands the potential part of Eq.\,(\ref{9}) to be filtered-out. Equation (\ref{9}) is curled for this purpose. The $z$-component of the resulting equation governs the toroidal flow. The poloidal flow equation is the $z$-component of the motion equation curled twice. Neglecting disturbances in pressure to exclude magnetic buoyancy instability (Acheson \& Gibbons 1978), the gravity term can be transformed as follows
\begin{eqnarray}
    \hat{\vec z}\cdot\left({\vec\nabla}\times\left({\vec\nabla}\times\rho{\vec g}\right)\right) &=&
    -\frac{\rho g}{c_\mathrm{p}}\left(\hat{\vec z}\times{\vec\nabla}\right)
    \cdot\left(\hat{\vec z}\times{\vec\nabla}\right)S
    \nonumber \\
    &=& -\frac{\rho g}{c_\mathrm{p}}\Delta_2 S,
    \label{15}
\end{eqnarray}
where $\Delta_2 = \partial^2/\partial x^2 + \partial^2/\partial y^2$ is the 2D Laplacian.
This leads to the following equation for the poloidal flow:
\begin{eqnarray}
    &&\frac{\partial}{\partial t}
    \left(\left(
    \frac{\partial^2}{\partial z^2} - k^2
    \right) V\right)\ =\ -\frac{\rho g}{c_\mathrm{p}} S
    \nonumber \\
    &&+\ \frac{\mathrm{i}k_2}{4\pi}\left(
    B\frac{\partial^2 P'}{\partial z^2} -
    \frac{\partial^2 B}{\partial z^2}P' - k^2 B P'
    \right)
    \nonumber \\[0.1 truecm]
    &&+\
    \left(
    \frac{\partial^2}{\partial z^2} - k^2
    \right)
    \left(
    \rho\nu_{_\mathrm{T}}\frac{\partial}{\partial z}\frac{1}{\rho}\frac{\partial V}{\partial z}
    - \nu_{_\mathrm{T}} k^2 V
    \right)
    \nonumber \\[0.1 truecm]
    &&+\ 2 k^2
    \left(
    \frac{1}{\rho}\frac{\partial^2 (\rho\nu_{_\mathrm{T}})}{\partial z^2}
    - \frac{1}{\rho^2}\frac{\partial (\rho\nu_{_\mathrm{T}})}{\partial z}
    \frac{\partial\rho}{\partial z}
    \right)
    V .
    \label{16}
\end{eqnarray}
Derivation of other equations of the full system does  not pre\-sent difficulties and requires no comment. These equations are therefore omitted.

Conditions at the bottom boundary assume an interface with a superconductor beneath, zero surface stress, zero normal components of the magnetic field and velocity, and the absence of
disturbances in the heat flux,
\begin{eqnarray}
    &&\frac{\partial}{\partial z}(\sqrt{\eta_{_\mathrm{T}}} T') =
    \frac{\partial}{\partial z}\left(\frac{W}{\rho}\right) =
    \frac{\partial S}{\partial z} = P' = V = 0,
    \nonumber \\
    &&(z = 0).
    \label{17}
\end{eqnarray}
All the disturbances were put to zero on the (artificial) surface boundary at $z = h$.

Equations for the disturbances were solved numerically with finite-difference representation of the derivatives in $z$. Inhomogeneity of solutions near the bottom boundary can be sharp. A non-uniform grid with higher density of grid-points near the bottom was, therefore, applied:
\begin{equation}
    z_1 = 0,\ z_i = h\left( 1 - \cos\left( \pi\frac{i - 3/2}{2N - 3}\right)\right),\ 2\leq i \leq N ,
    \label{18}
\end{equation}
where $N$ is the grid-point number. The results to follow were obtained with $N = 52$. Trial computations with larger $N$ gave practically the same results.

Searching for the dependence of small disturbances on time in the exponential form,
$\mathrm{exp}(\sigma t)$, leads to the eigenvalue problem (from now on, $\sigma$ is the eigenvalue of the linear stability problem). A positive real part $\Re(\sigma) > 0$ means an instability.

Obviously, an instability can emerge even without the magnetic field if too small turbulent transport coefficients are prescribed (Tuominen et al. 1994). The smaller
the turbulent diffusivity, the larger the entropy gradient (superadiabaticity) in
the background state. The usual convective instability develops for sufficiently
small diffusivity. The problem formulation then loses consistency because the turbulent transport coefficients no longer parameterise the convective turbulence adequately. The value of the turbulent diffusivity (\ref{7}) is controlled by the
$\alpha_{_\mathrm{MLT}}$-parameter.

\begin{figure}[!t]
\includegraphics[width=8 truecm]{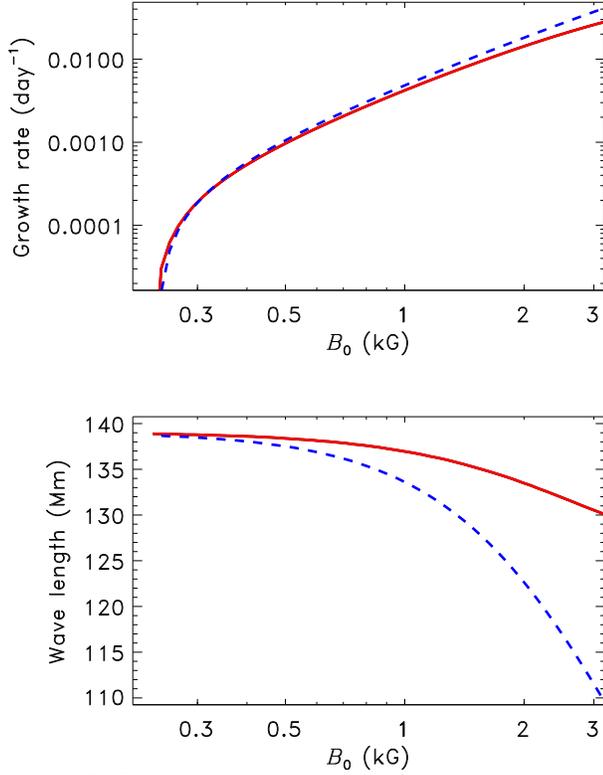}
 \caption{{\sl Top panel:} growth rates of bending (full line) and
        interchange (dashed) modes in dependence on $B_0$. {\sl Bottom panel:} The
        wave lengths of bending and interchange disturbances corresponding to the maximum growth rates of the upper panel.
    }
 \label{f3}
\end{figure}

$\alpha_{_\mathrm{MLT}} = 0.48$ is the marginal value for the onset of convective instability. The slightly supercritical value $\alpha_{_\mathrm{MLT}} = 0.49$ is used in the computations to follow. Such a choice was justified in a preceding paper (Kitchatinov \& Mazur 2000). The relatively small value of $\alpha_{_\mathrm{MLT}} = 0.49$ is related to the consideration of the deep near-bottom region of the convection zone. For higher regions, the critical values of $\alpha_{_\mathrm{MLT}}$ are larger. $\alpha_{_\mathrm{MLT}}$ should probably decrease with depth in a more realistic model.

It can be seen from the equations for small perturbations that the eigenvalues do not change with an inversion of sign of the wave number $k_1$ or $k_2$ or both. We therefore consider the positive wave numbers only.
\section{Results and discussion}
Eigenvalues of all unstable modes are real (change of stability). In no case do the largest growth rates belong to the modes whose components $k_1$ and $k_2$ of the wave vector differ from zero simultaneously. Depending on the value of $B_0$, the disturbances with the wave vector along either the $x$ or $y$ axis show the most rapid growth. Only these two cases are therefore discussed. The disturbances with $k_1 =0$ and $k_2 \neq 0$ deform (bend) the field lines. These disturbances will be called \lq bending modes'. In the case of $k_1 \neq 0$ and $k_2 =0$ the field lines are interchanged without bending. Such disturbances will be called \lq interchange modes'.

The magnetic field can oppose flow bending the field lines. It can therefore be expected that magnetic quenching of turbulent diffusivity leads to an instability of interchange modes which do not bend the lines. Figure 3, however, shows close growth rates for bending and interchange modes. For weak fields, bending modes grow even faster. This is a consequence of the magnetic suppression of the thermal diffusivity. If the suppression is neglected (the first term on the right-hand side of Eq.\,(\ref{14}) is dropped), interchange mode dominates for any $B_0$.

\begin{figure}[]
\includegraphics[width=8 truecm]{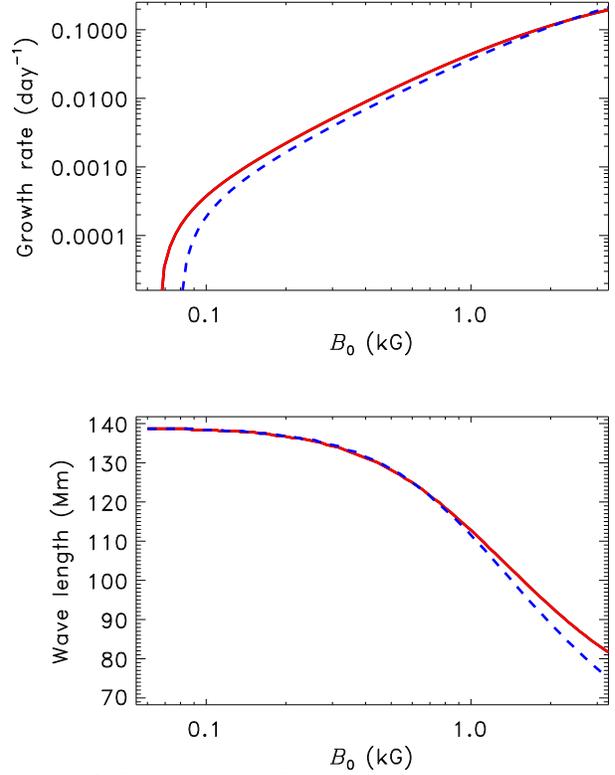}
 \caption{The same as in Fig.\,3 but with $B_\mathrm{eq}$
         decreased three times compared to Fig.\,1.
    }
 \label{f4}
\end{figure}

Numerical experiments by Karak et al. (2014) have shown that equations (\ref{5}) and (\ref{6}) reproduce satisfactorily the diffusivity quenching but only if $B_\mathrm{eq}$ is defined not for the original turbulence, which would take place in the absence of the magnetic field, but for the actual magnetized flow. This is equivalent to a reduced value of $B_\mathrm{eq}$ compared to our estimations. The computations were repeated with $B_\mathrm{eq}$ reduced three times compared to Fig.\,1. The results of these computations are shown in Fig.\,4. The bending modes are dominating now in a wider range of $B_0$ values and the growth rates increase considerably. The difference between thermal and magnetic energies also increases compared to Fig.\,2. The structure of unstable disturbances, which will be discussed shortly, depends weakly on the definition of $B_\mathrm{eq}$. The discussion to follow refers to the definition of $B_\mathrm{eq}$ of the Eq.\,(\ref{7}) and Fig.\,1.

\begin{figure}[!t]
\includegraphics[width=8 truecm]{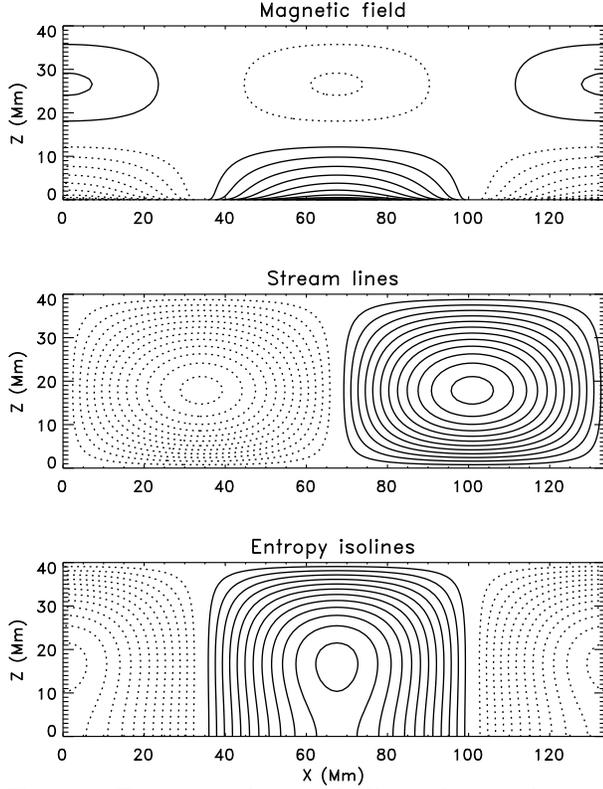}
 \caption{{\sl From top to bottom:} isolines of magnetic disturbances,
        stream-lines of the flow, and the entropy disturbances isolines for the most rapidly growing interchange mode for $B_0 =1000$\,G. The full (dashed) lines show positive (negative) levels and clockwise (anti-clockwise) circulation.
    }
 \label{f5}
\end{figure}

Figure~5 shows the structure of the most rapidly growing interchange mode for $B_0 = 1$\,kG. The magnetic field disturbances are concentrated in the lower part of the layer. The field amplifications (positive disturbances) are connected with the converging horizontal flows. The upward flows rise from relatively hot regions of positive entropy disturbances.

\begin{figure}[!t]
\includegraphics[width=8 truecm]{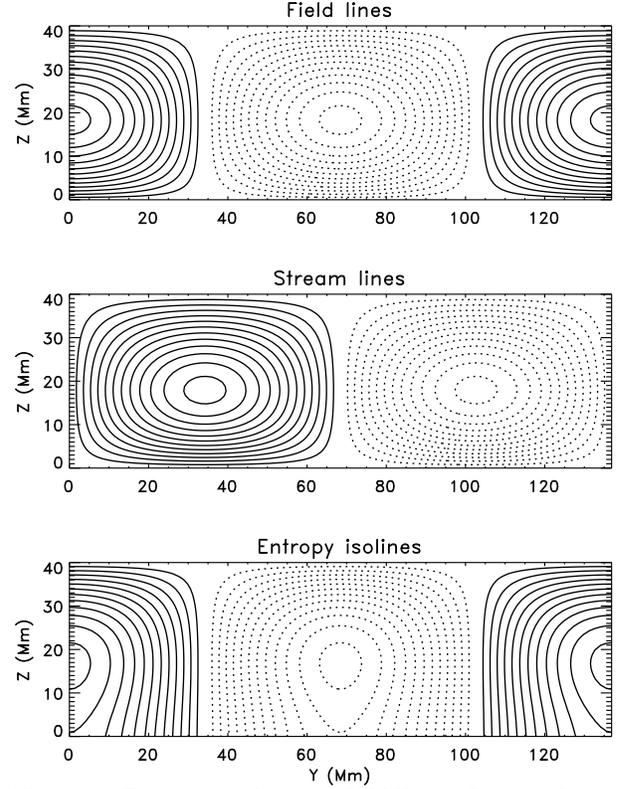}
 \caption{{\sl From top to bottom:} field lines of magnetic
            disturbances, stream-lines of the flow, and isolines of the entropy
            disturbances of the bending mode for $B_0 =1000$\,G. The full
            (dashed) lines show positive (negative) levels and clockwise
            (anti-clockwise) circulation.
    }
 \label{f6}
\end{figure}

The most rapidly growing bending mode is shown in Fig.\,6. The field structure resulting from this unstable mode is shown in Fig.\,7 as a superposition of the background field and its disturbance with an amplitude of about 30\% of the background field. In contrast with the interchange mode of Fig.\,5, inhomogeneity along the $y$-direction of the background field is now present. Horizontal flows along the background field lines do not disturb the field. Field amplification near the base of the layer occurs in the regions of downflow.

\begin{figure}[]\centering
\includegraphics[width=\linewidth]{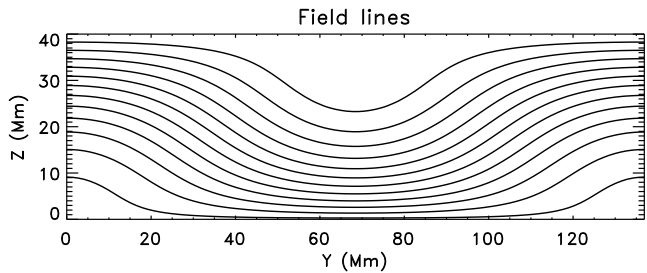}
 \caption{Superposition of the background magnetic field and
        magnetic disturbances of the bending mode of Fig.\,6. Amplitude of the disturbance is about 30\% of the background field.
    }
 \label{f7}
\end{figure}

An interchange of the field lines is unlikely to produce a considerable field amplification. The bending mode of Figs.\,6 and 7 can be more relevant to the hypothetical formation of strong field regions. In this case, redistribution of the fluid along the field lines can change the field strength considerably. However, the bending disturbances do not produce flux-tubes: they are homogeneous along the $x$-axis normal to the background field. Increased field regions of finite dimensions in both horizontal directions can result from a superposition of the bending and interchange disturbances (similar to the B\'ernard cells formation by a superposition of the linear modes of thermal convection with different directions of horizontal wave vectors; Chandrasekhar 1961, p.47).

Comparison of the magnetic and thermal energies of Fig.\,2 suggests that the field amplification in unstable disturbances can be considerable. The amount of the amplification can be evaluated with nonlinear computations only. Some order-of-magnitude estimations are nevertheless possible from the linear computations. Figures 5 and 6 show the characteristic scales of the field amplification regions in horizontal dimensions $L_x \approx L_y \approx 50$~Mm. The disturbances are localized near the base of the convection zone. Their vertical size $L_z \approx 10$~Mm. The magnetic energy $W_\mathrm{M} \approx B^2 L_x^2 L_z/(8\pi)$ in this region can be estimated as
\begin{equation}
    W_\mathrm{M} \approx 10^{33} \left(\frac{B}{1\mathrm{kG}}\right)^2\ \mathrm{erg} ,
    \label{19}
\end{equation}
where $B$ is the characteristic strength of the background field. The rough estimation (\ref{19}) seems to be the first attempt at connecting the large-scale fields of the solar dynamo with parameters of the active regions. The estimation however agrees in order of magnitude with the magnetic energy of the active regions (Sun et al. 2012; Livshits et al. 2015).

Dynamo models give toroidal fields of several kilo-Gauss near the base of the solar convection zone. This field should be amplified tens times by the presumed instability for its rise to the solar surface to fit the observational properties of the active regions (D'Silva \& Choudhuri 1993). The field strength in sunspots suggests that the flux-tube expansion in the course of the rise reduces the field strength again to several kilo-Gauss keeping the estimation (\ref{19}) for the magnetic energy. Figures 3 and 4 show that the characteristic time of several months for the instability is short compared to the 11-year period of the solar cycle.
\phantomsection
\section*{Acknowledgments}
This work was supported by budgetary funding of Basic Research program II.16 and by the Russian Foundation for Basic Research (project  17-02-00016).
\phantomsection
\section*{References}
\begin{description}
\item Acheson,~D.\,J., \& Gibbons,~M.\,P. 1978, Phyl. Trans. Roy. Soc. London {\bf A289}, 459
\item Babcock,~H.\,W. 1961, \apj\ {\bf 133}, 572
\item Caligari,~P., Moreno-Insertis,~F., \& Sch\"ussler,~M. 1995, \apj\ {\bf 441}, 886
\newpage
\item Chandrasekhar,~S. 1961, Hydrodynamic and Hydromagnetic Stability (Oxford: Clarendon Press)
\item Dasi-Espuig,~M., Solanki,~S.\,K., Krivova,~N.\,A., Cameron,~R., \& Pe\~{n}uela,~T. 2010, \aap\ {\bf 518}, A7
\item D'Silva,~S., \& Choudhuri,~A.\,R. 1993, \aap\ {\bf 272}, 621
\item Erofeev,~D.\,V. 2004, in: Multi-Wavelength Investigation of Solar Activity, IAU Symp. 223 (Eds. A.V.\,Stepanov, E.E.\,Benevolenskaya, A.G.\,Kosovichev, Cambridge, \\ UK: Cambridge Univ. Press), p.97
\item Gilman,~P.\,A. 1986, in: Physics of the Sun (Ed. P.A.\,Sturrock, Dordrecht, D. Reidel Publishing Company), Vol.1, p.95
\item Gilman,~P.\,A., \& Glatzmaier,~G.\,A. 1981, \apjs\ {\bf 45}, 335
\item Hale,~G.E., Ellerman,~F., Nicholson,~S.B., \& Joy,~A.\,H. 1919, \apj\ {\bf 49}, 153
\item Karak,~B.\,B., Rheinhardt,~M., Brandenburg,~A., K\"apyl\"a,~P.\,J., \& K\"apyl\"a,~M.\,J. 2014, \apj\ {\bf 795}, 16
\item Khlystova,~A., \& Toriumi,~S. 2017, \apj\ {\bf 839}, 63
\item Kitchatinov,~L.\,L., \& Mazur,~M.\,V. 2000, \sp\ {\bf 191}, 325
\item Kit\-cha\-ti\-nov,~L.\,L., \& Olemskoi,~S.\,V. 2006, Astron. Lett. {\bf 32}, 320
\item Kit\-cha\-ti\-nov,~L.\,L., \& Olemskoy,~S.\,V. 2011, Astron. Lett. {\bf 37}, 656
\item Kitchatinov,~L.\,L., Pipin,~V.\,V., \& R\"udiger,~G. 1994, \an\ {\bf 315}, 157
\item Krause,~F., \& R\"adler,~K.-H. 1980, Mean-Field Magnetohydrodynamics and Dynamo Theory (Berlin, Akademie-Verlag)
\item Leighton,~R.\,B. 1969, \apj\ {\bf 156}, 1
\item Lites,~B.\,W., Skumanich,~A., \& Martinez\,Pillet,~V. 1998, \aap\ {\bf 333}, 1053
\item Livshits,~M.\,A., Rudenko,~G.\,V., Katsova,~M.\,M., \& Mysh\-ya\-kov,~I.\,I. 2015, Adv. Space Res. {\bf 55}, 920
\item Parker,~E.\,N. 1979, Cosmical Magnetic Fields (Oxford, Clarendon Press)
\item Parker,~E.\,N. 1984, \apj\ {\bf 283}, 343
\item Stix,~M. 1989, The Sun (Springer, Berlin)
\item Sun,~X., Hoeksema,~J.\,T., Liu,~Y., Wiegelmann,~T., Hayashi,~K., Chen,~Q., \& Thalmann,~J. 2012, \apj\ {\bf 748}, 77
\item Tuominen,~I., Brandenburg,~A., Moss,~D., \& Rieutord,~M. 1994, \aap\ {\bf 284}, 259
\item Weber,~M.\,A., Fan,~Y., \& Miesch,~M.\,S. 2011, \apj\ {\bf 741}, 11
\item Weiss,~N.\,O. 1966, Proc. Roy. Soc. London {\bf A293}, 310

\item Yousef,~T.\,A., Brandenburg,~A., \& R\"udiger,~G. 2003, \aap\ {\bf 411}, 321
\item Zel'dovich,~Ya.\,B. 1957, Sov. Phys. JETP {\bf 4}, 460
\item Zwaan,~C. 1992, in: Sunspots: Theory and Observations (Eds. J.H.\,Thomas, N.O.\,Weiss, Kluwer Academic Publishers), p.75
\end{description}
\end{document}